\documentclass[final,1p,times]{elsarticle} 
\usepackage{graphicx} 
\usepackage{amssymb} 
\usepackage{amsthm} 
\usepackage{lineno} 

\journal{Nuclear Physics A} 
\begin{document} 

\begin{frontmatter} 


\title{Single Particle Probes of d+Au Collisions in PHENIX}

\author{Zvi Citron for the PHENIX collaboration}

\address{Stony Brook University, 
Department of Physics and Astronomy,
Stony Brook, NY, 11794, USA}

\begin{abstract} 
Deuteron-gold collisions provide insights into the nuclear structure
function and a valuable baseline for Au+Au collisions.  Measurement
of the nuclear modification factor, $R_{dAu}$, in d+Au in the PHENIX
central arms for hadrons and photons allows us to disentangle
cold nuclear matter effects from the hot medium effects that are
important in Au+Au collisions.  In addition, the d+Au system
can yield important insights into the gluonic structure of the Au nucleus.
RHIC experiments have previously measured suppression of forward
rapidity particle production relative to p+p scaled by the number of
binary N-N collisions, but a definitive explanation of these data is thus
far elusive.  Correlations between hadrons with a large rapidity gap are
a particularly sensitive probe of gluon saturation.  We will discuss probing
this physics via particle production in events tagged with high momentum
particles at different rapidities, along with $R_{dAu}$ in new forward
rapidity regions.
\end{abstract} 

\end{frontmatter} 




The 2008 RHIC run of deutreron-gold collisions, with $\approx$ 30 times increase in statistics with respect to the 2003 dataset,  provided for important measurements of the properties of cold nuclear matter.  PHENIX has measured the mid-rapidity charged hadron production to calculate the nuclear modification factor, $R_{dAu}$, and has made use of new forward rapidity detectors to gain sensitivity to different regions in $x$. 

\begin{figure}[ht]
\centering
\includegraphics[width=1.0\textwidth]{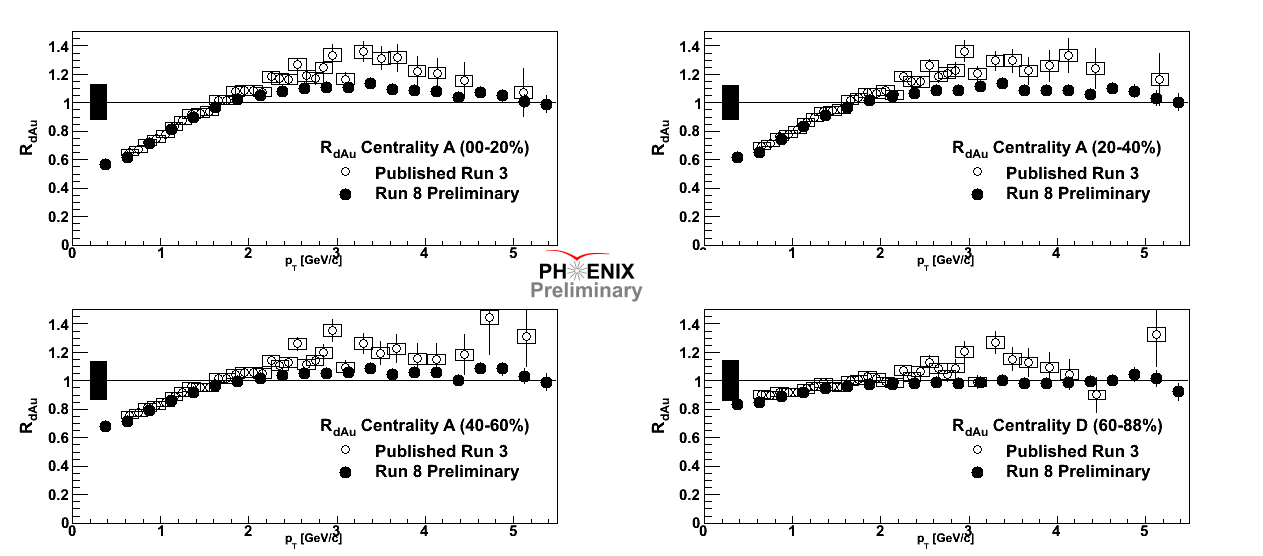}
\caption[]{Mid-rapidity inclusive charged hadron R$_{dAu}$ measured by PHENIX in the 2003 and 2008 RHIC runs.  The solid rectangle indicates global systematic uncertainty, this uncertainty may not fully cancel between the two data sets.
}
\label{rda_pic}
\end{figure}

The nuclear modification factor is defined as:
\begin{equation}
R_{dAu}(p_T) = \frac{(1/N_{evt}^{d+Au}) \; d^{2}N^{d+Au}/dp_T d\eta }
{\langle N_{coll}\rangle (1/N_{evt}^{p+p})\; d^{2}N^{p+p}/dp_T d\eta}. 
\label{eq:RAA_defined}
\end{equation} 
It is calculated separately for different centrality bins as categorized by the PHENIX Beam-Beam Counters (BBC), with each centrality bin yield appropriately corrected due to BBC effects\cite{ppg041}. This quantity is a useful baseline to gauge the impact of cold nuclear matter effects as compared to the analogous quantity $R_{AuAu}$ which shows suppression due to plasma effects \cite{AA_sup}.  It is also important for understanding other effects which are difficult to isolate in a heavy ion environment.  Figure \ref{rda_pic} shows the mid-rapidity PHENIX measurements from the 2003 RHIC run \cite{ppg041} as well as the new preliminary data from the higher statistics 2008 run.

Besides the nuclear modification factor measured at mid-rapidity, RHIC experiments have previously measured the nuclear modification factor in d+Au collisions at forward, d going side, and backward, Au going side, rapidities and found suppression in the forward direction \cite{ppg036,brahms}. A definitive explanation of this observation is thus
far elusive, but there are several models which seek to explain it.  To further study these phenomena PHENIX installed forward (North) and backward (South) electromagnetic calorimeters, named the Muon Piston Calorimeters (MPC) for their location in PHENIX, covering 3.1 to 3.9 and -3.1 to -3.7 in $\eta$ respectively.  Several of the models are sensitive to correlations between particles at forward rapidity and those at mid-rapidity \cite{vitev,dima}.  To study this we can trigger on particles in the forward (and backward) MPC to access relatively lower (and higher) $x$ and measure the inclusive mid-rapidity charged particle yield.  
To trigger in the MPC a $\pi^{0}$ is reconstructed from two electromagnetic calorimeter clusters.  At energy greater than $\approx$ 17 GeV a $\pi^{0}$ can not be reliably distinguished from a single $\gamma$ due to cluster overlap, so to  reach higher energy a separate data set is kept in which the triggers are simply inclusive electromagnetic clusters (the preponderance of $\gamma$s are from $\pi^{0}$ decays).  Although the two MPCs do not have identical acceptances to enable a direct comparison of the forward rapidity triggered to backward rapidity triggered data sets the acceptances are ``symmetrized".  An occupancy correction is applied to the measured trigger energy which is  $\approx \%$10 for triggers in the MPC on the Au going side in d+Au collisions.

\begin{figure}[ht]
\centering
\includegraphics[width=0.65\textwidth]{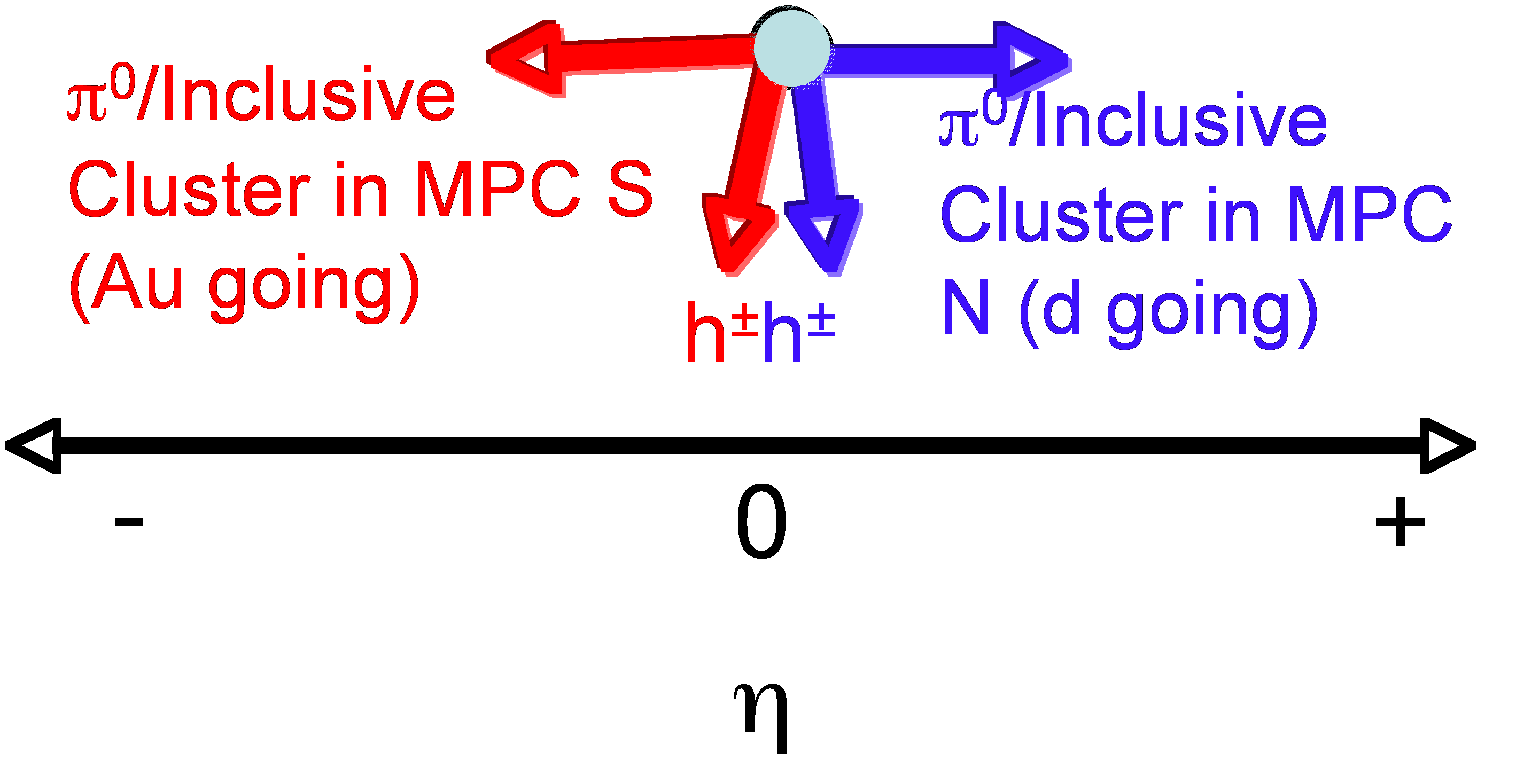}
\caption[]{A cartoon of the measurement.  Both forward (N) and backward (S) rapidity triggered type events are illustrated.}
\label{schem}
\end{figure}

\begin{figure}[ht]
\centering
\includegraphics[width=0.45\textwidth]{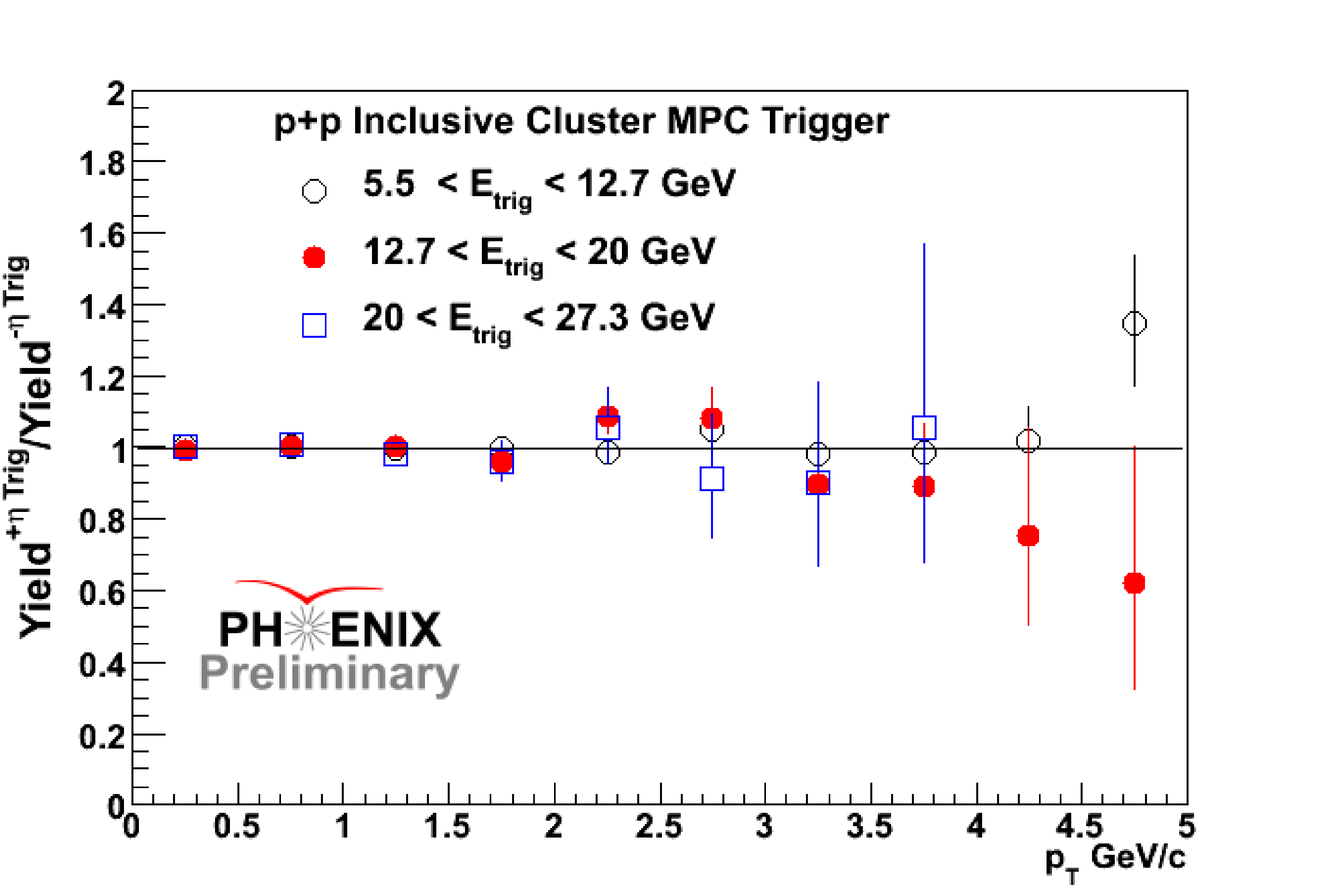}
\includegraphics[width=0.45\textwidth]{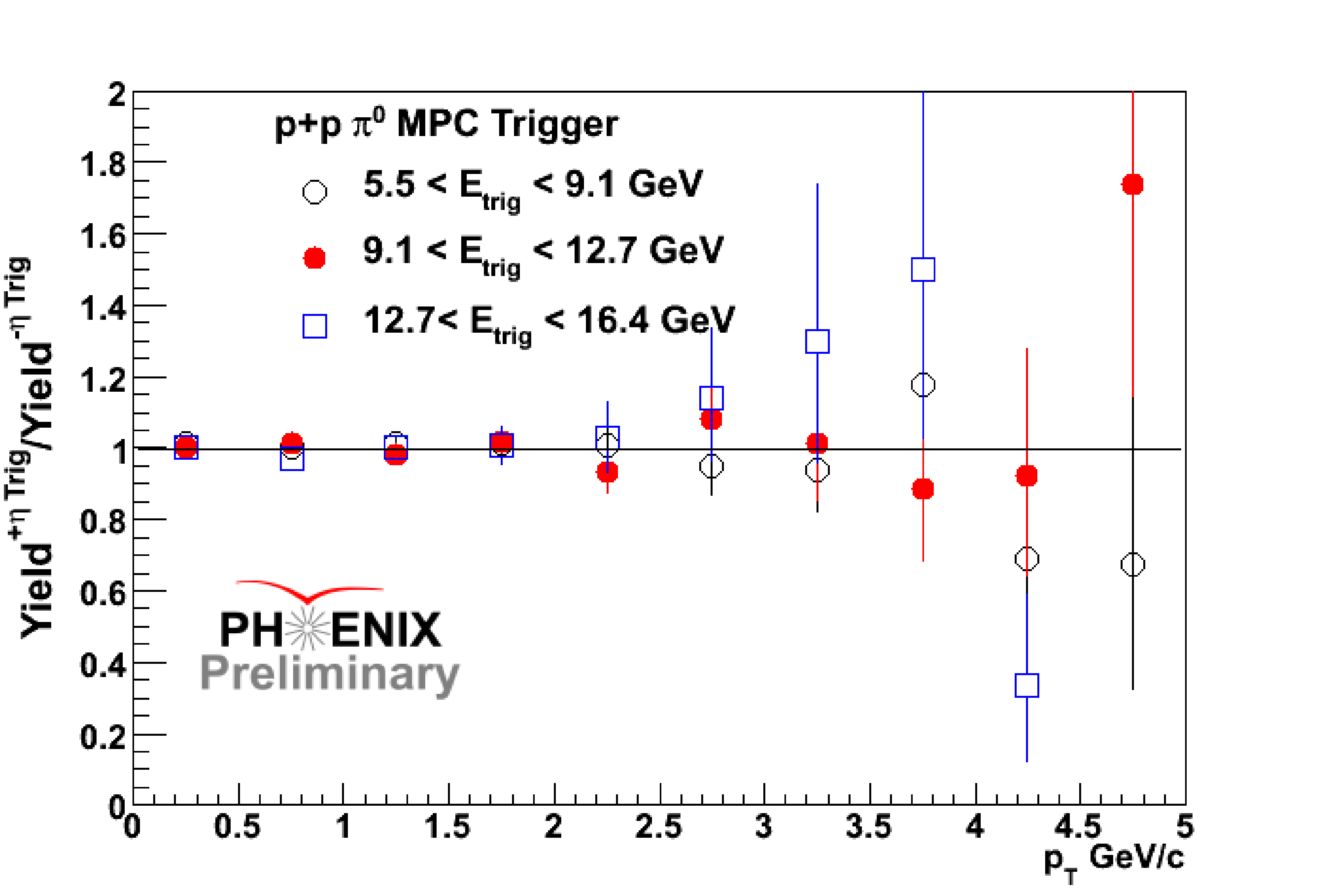}
\caption[]{The ratios of the forward MPC triggered mid-rapidity h$^{\pm}$ to the backward triggered mid-rapidity h$^{\pm}$ for different trigger energies in p+p collisions.  In the left plot the MPC trigger is an inclusive EM cluster, and on the right the MPC trigger is a reconstructed $\pi^{0}$.}
\label{pp_rats}
\end{figure}

With the triggers thus defined, the mid-rapidity inclusive charged hadron per trigger yield can be measured for forward and backward rapidity triggers. 
A cartoon schematic of the measurement is shown in figure \ref{schem}. 
In the symmetric p+p system the labels of forward and backward rapidity are arbitrary and we expect that the mid-rapidity per trigger yield will be the same regardless of whether the trigger is ``forward" or ``backward".  To show that this is indeed the case the ratio of the mid-rapidity yields of the two trigger types, R$_{FB}$ is defined:
\begin{equation}
R_{FB}(p_T) = 
\frac{(1/N_{3.1<\eta<3.7}) \; d^{2}N_{|\eta| < 0.35}^{+\eta \ trig}/dp_T d\eta }
{(1/N_{-3.1>\eta>-3.7}) \; d^{2}N_{|\eta| < 0.35}^{-\eta \ trig}/dp_T d\eta}. 
\label{eq:RFB_defined}
\end{equation} 
R$_{FB}$ of p+p collisions for three different trigger energies is plotted in figure \ref{pp_rats}.  It is consistent with unity for all triggers, as expected, demonstrating that there is no lurking artificial asymmetry due to detector effects.

\begin{figure}[!ht]
\centering
\includegraphics[width=0.62\textwidth]{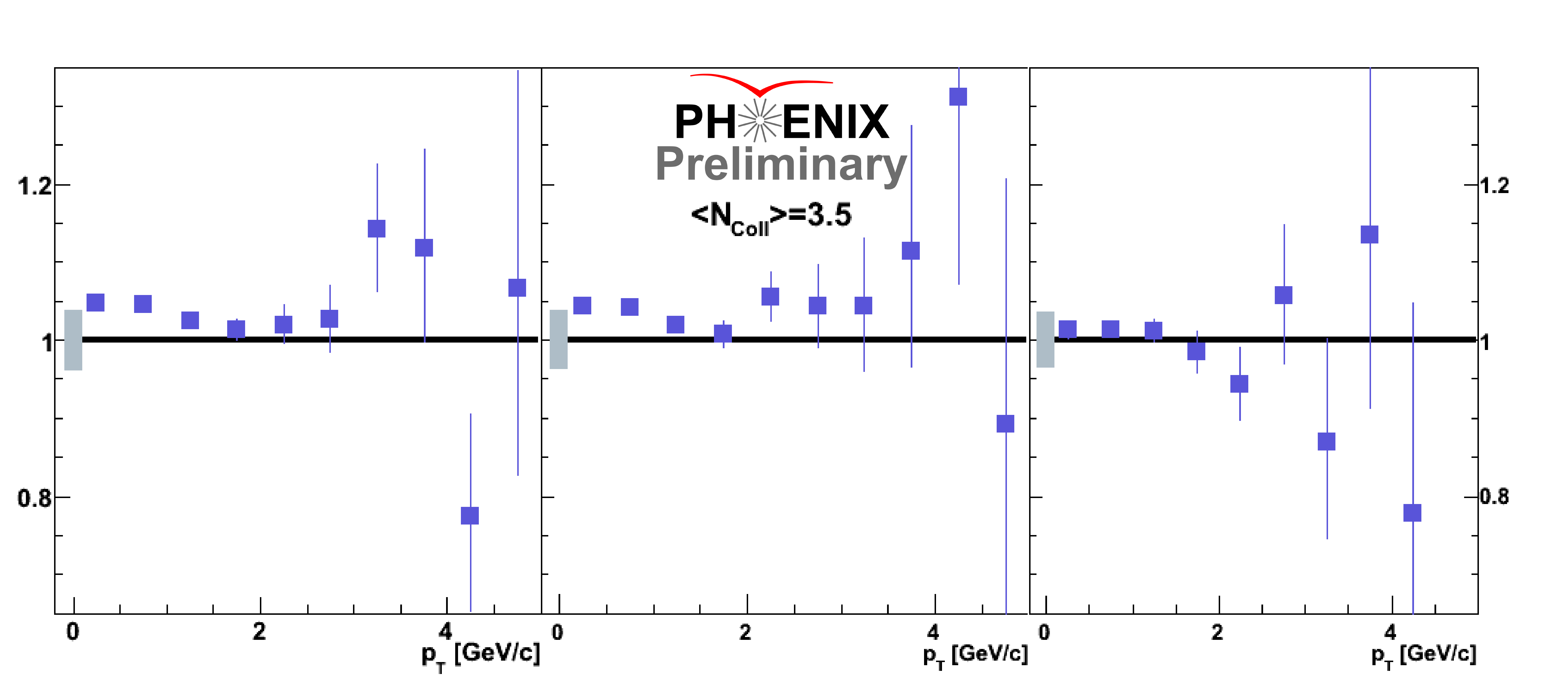}
\includegraphics[width=0.62\textwidth]{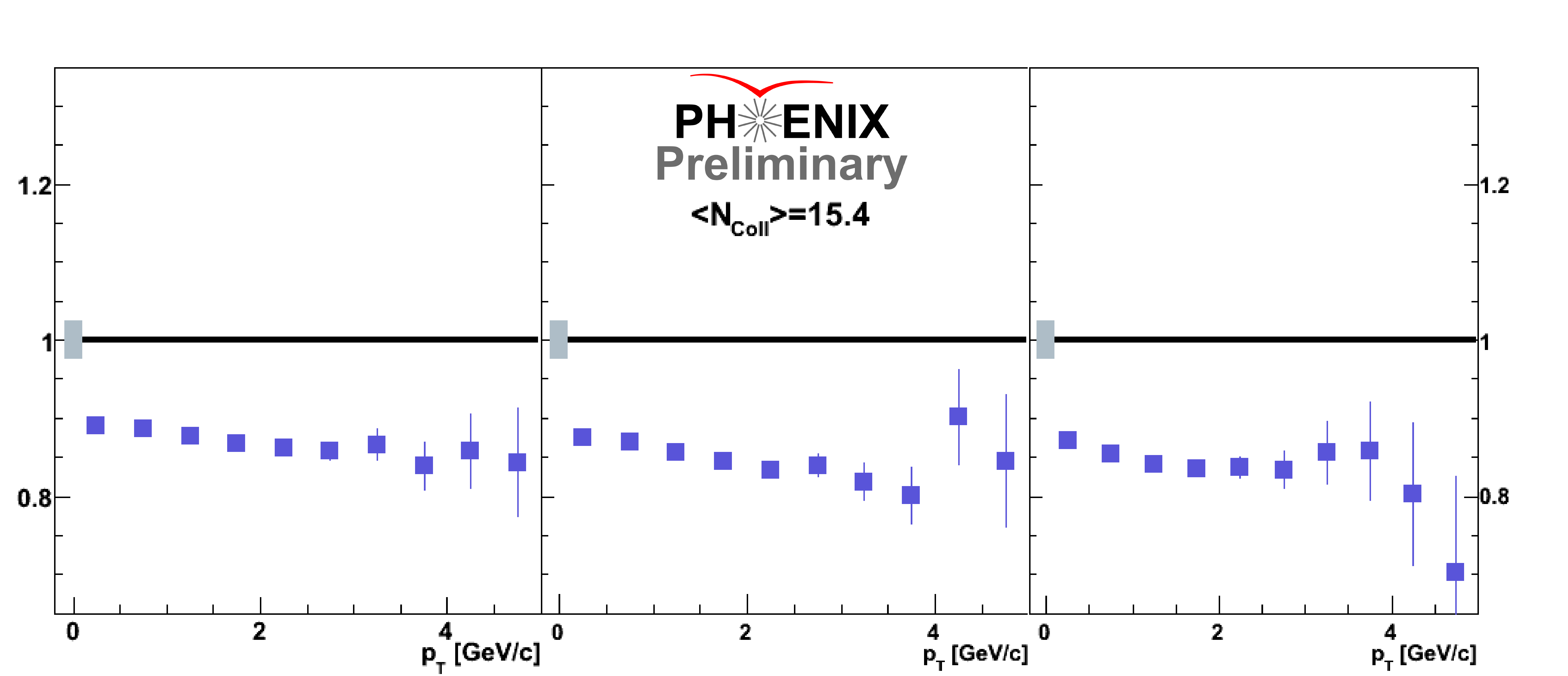}
\caption[]{The ratio of the d going side $\pi^{0}$ MPC triggered mid-rapidity h$^{\pm}$ to the Au going side $\pi^{0}$ triggered mid-rapidity h$^{\pm}$ in the most peripheral collisions ($<N_{coll}>$ = 3.5) on top and the most central collisions ($<N_{coll}>$ = 15.4) on the bottom.  From left to right the three panels are for $\pi^{0}$ trigger energies of 5.5$<$E$<$9.1, 9.1$<$E$<$12.7, and 12.7$<$E$<$16.4 GeV.  Bands indicate overall systematic uncertainties.}
\label{samp_rat}
\end{figure}

\begin{figure}[!ht]
\centering
\includegraphics[width=0.48\textwidth]{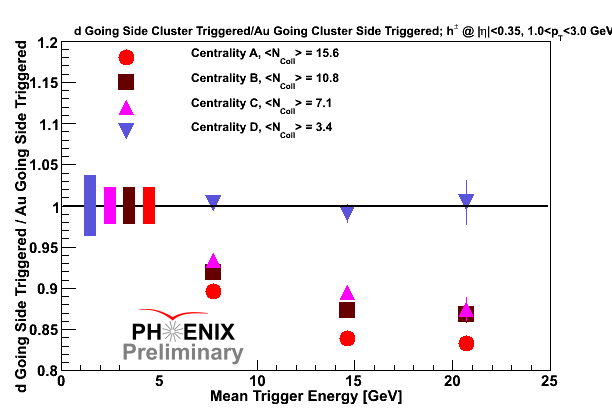}
\includegraphics[width=0.48\textwidth]{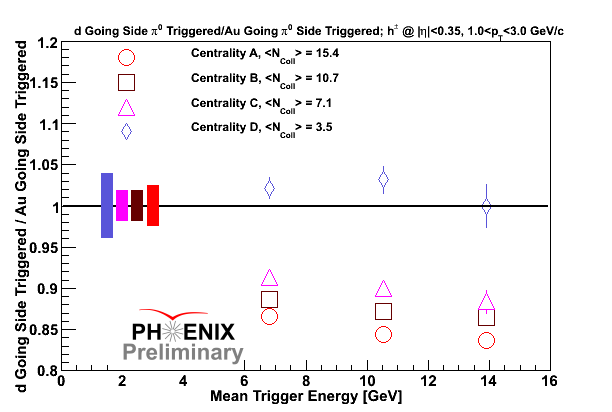}
\caption[]{The suppression of the d going side triggered yield relative to the Au going side triggered yield as a function of the trigger energy.  The four bands are for the four centrality selections (from left to right least to most central) and indicate systematic uncertainties from the centrality bias associated with the trigger requirement and possible asymmetries stemming from the higher multiplicity in the Au going side MPC.
}
\label{summary}
\end{figure}

Of more interest than the p+p system is the d+Au system, in which forward and backward rapidity are not arbitrary but refer to the d going and Au going side, respectively.  In the d+Au system shadowing, saturation, or other effects may lead to R$_{FB}$ $\neq$ 1.  R$_{FB}$ is calculated for four centrality bins to assess the significance of the nuclear volume at play in the collision.  In the most peripheral collisions ($<N_{coll}>$ = 3.5) we expect results similar to those in p+p as they are shown to be in the top panel of figure \ref{samp_rat}.  However, in more central collisions the d going side triggered mid-rapidity inclusive charged hadron spectra are suppressed compared to the Au going side triggered sample.  This is seen in the bottom panel of  figure \ref{samp_rat} where R$_{FB}(p_{T})$ $<$ 1.  Figure \ref{summary} summarizes the results for different triggers in different centrality bins, by showing R$_{FB}$ integrated over 1.0 $<p_{T}<3.0$ GeV/c as a function of the trigger energy.

Further study is necessary to contextualize this study in terms of an understanding of the physics underlying the observations, in particular, whether these data support shadowing, anti-shadowing, or saturation in the nucleus.  See also a complementary analysis of the same data set in these proceedings \cite{beau}.

\clearpage


\end{document}